# Identification and Characterization for Disruptions in the U.S. National Airspace System (NAS)

Jing Xu, Mark Hansen
Department of Civil and Environmental Engineering
University of California, Berkeley
Berkeley, United States
jing-xu@berkeley.edu
mhansen@ce.berkeley.edu

Megan S. Ryerson
Department of City and Regional Planning
Department of Electrical and Systems Engineering
University of Pennsylvania
Philadelphia, United States
mryerson@design.upenn.edu

*Abstract*—Disruptions in the National Airspace System (NAS) lead to significant losses to air traffic system participants and raise public concerns. We apply two methods, cluster analysis and anomaly detection models, to identify operational disruptions with geographical patterns in the NAS since 2010. We identify four types and twelve categories of days of operations, distinguished according to air traffic system operational performance and geographical patterns of disruptions. Two clusters--NAS Disruption and East Super Disruption, accounting for 0.8% and 1.2% of the days respectively, represent the most disrupted days of operations in U.S. air traffic system. Another 16.5% of days feature less severe but still significant disruptions focused on certain regions of the NAS, while on the remaining 81.5% of days the NAS operates relatively smoothly. Anomaly detection results show good agreement with cluster results and further distinguish days in the same cluster by severity of disruptions. Results show an increasing trend in frequency of disruptions especially post-COVID. Additionally, disruptions happen most frequently in the summer and winter.

## I. Introduction

Disruptions in the U.S. National Airspace System (NAS) lead to significant losses of resources, raise public concern on the reliability of the aviation system, and degrade intercity mobility and accessibility. Yet, despite extensive experiences with disruptions, "playbooks" regarding disruptions focus on emergency weather routes rather than the coordinated system recovery required to preserve mobility of passengers and goods. To this end, this study endeavors to make a fundamental contribution: to develop a rigorous methodology to define typologies of disruptions in the NAS. This methodology will serve as the foundation for research on countermeasures to disruptions; it will allow us to understand the causes of disruptions and identify mitigation options ultimately increasing the resilience of the NAS in the future.

Our approach is to develop two models, a cluster analysis model and an anomaly detection model, to identify disruptions in the NAS back to 2010. We perform both analyses on operational variables and then characterize their geographical patterns. Two different methods enable us to compare the agreements and disagreements on identified results; in short, it's a method to cross-validate the methods. The long analysis period provides opportunities to investigate if there are any trends in disruptions since 2010.

The scholarship in this space, to date, has focused on specific disruptions: the cause of a disruption, the recovery processes, and the development of strategies to deal with similar issues in the future. Sun et al. [1] point out the extreme vulnerability of our aviation system towards external disruptions and analyze the full disruption cycle of COVID with a three-stage disruption and recovery story: the epidemic shock, over the pandemic stalemate, and towards the endemic-induced recovery. Although some suggestions for constructing a pandemic-resilient aviation system are discussed in this study, it does not provide further detailed strategies that airlines or authorities can directly employ. Motivated by impacts of severe weather on air traffic system, Glass et al. [2] develop methods to identify air traffic disruptions caused by weather, and a binary integer programming optimization model to generate schedules and minimize delays and help recover from disruptions with a case study for Hurricane Matthew. This paper focuses on flight schedules, while methods will be optimized with consideration of other components of air traffic management, like crew assignment.

Our study builds on existing studies that seek to identify disruptions over geographically limited areas of the NAS using cluster analysis. Kuhn [3] applies cluster analysis to air traffic data and identifies 5 similar types of days for the New York area, which help to facility air traffic management in the area. Gorripaty et al. [4] conduct similar cluster model and identify similar days of operations at four case study airports – EWR, SFO, ORD, and JFK. Rebollo and Balakrishnan [5] look into the top 112 airports and 584 OD pairs out of 2,029 airports and 31,905 OD pairs in the NAS and find 6 representative clusters with different operation patterns for year 2007 and 2008. These studies identify operational patterns for part of the NAS. For disruptions in the entire NAS, Xu et al. [6] compare aggregated NAS operational performance pattern pre- and post-COVID and find that NAS performance hasn't recovered from COVID as of summer 2023.

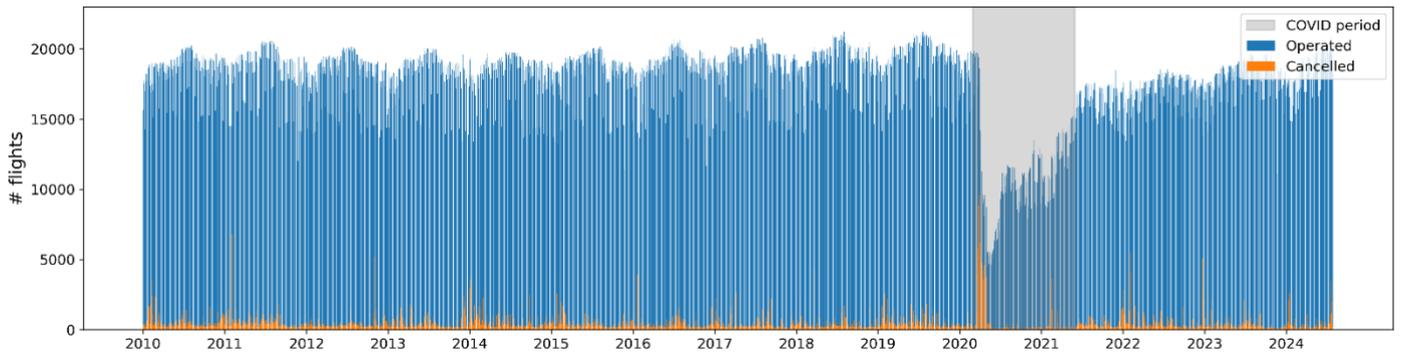
Figure 1. Daily Number of Operated and Cancelled Flights in Analysis Period

Our approach broadens the existing approaches by geographically encompassing the entire NAS and identifying typologies of disruptions with operational variables and geographical scope. This allows us to identify multi-regional disruptions, regional disruptions, and more localized disruptions with the same methodology.

Methods to identify disruptions could help scholars and aviation practitioners identify countermeasures to the wide-ranging impacts of disruption. Consider that the following events caused widespread disruption, with airlines and the entire system lacking a coordinated strategy for response. During the CrowdStrike outage in the summer of 2024, cancellations and delays soared; almost every major airline had immediate and cascading issues related to rescheduling, crew availability, and customer service which took several days to recover [7]. Following a major snowstorm in December 2022, Southwest Airlines had a scheduling crisis for 10 days following; their IT system was unable to keep pace with where their crews and planes were located. Southwest Airlines was fined $140 millions and paid over $600 million to compensate disrupted passengers [8]. Other examples abound, including the grounding of the Boeing 737 MAX in 2024 [9] and the January 2023 pause of all operations due to the outage of the NOTAM system [10]. During these periods, the disruptions in aviation system led to increased interest in developing ground-based substitutions to compensate during several circumstances [11, 12].

The remainder of this paper is organized as follows: in §II we discuss data sources and available variables; in §III we present the disruption identification methods and feature engineering; in §IV we present the analysis results and discuss their implications; and in §V we conclude this study and discuss the future work.

## II. DATA SOURCES

In this study, our research goal is to identify typologies of disrupted days of operations based on delays, cancellations, and geographical patterns thereof. We focus on days of operations for scheduled commercial flights in the NAS since 2010 excluding the COVID period. The precise analysis period is from January 1st, 2010, to July 31st, 2024, excluding March 1st, 2020, to June 30th, 2021, a total of 4,869 days.

We collect data from the Aviation System Performance Metrics (ASPM) dataset from FAA Operations & Performance Data [13]. ASPM dataset contains multiple subsets and documents flights to and from the ASPM airports (including the Core 30 and OEP 35 airports), and all flights by ASPM carriers.

ASPM Individual Flight Dataset records operational information for operated individual flights; almost all scheduled commercial flights in the NAS are included in the individual flight dataset. For each flight, the scheduled origin and destination airports, operational scheduled times, flight plan times, and actual times (including departure time, wheels-off time, wheels-on time, and arrival time) are all documented in this dataset. ASPM Derived Cancelled Flights Dataset records scheduled flight information for cancelled flights.

The number of operated flights and cancelled flights for each day of operations are visualized in Figure 1. The figure shows both the seasonality of operated and cancelled flights as well as indicates some severe outlier days with high cancellations and reduced operations. It is these days our methodology intends to identify and categorize into typologies.

## III. METHODOLOGY

The methodology seeks to identify typologies of disrupted days. Determination of the unit of analysis must be done carefully: should we analyze operational and delay metrics at the airport level, we would lose the regional characteristics and geographic nature of disruptions. We therefore first define airport groups to capture regional characteristics. We then perform feature engineering to select representative features to describe daily performance for each geographical airport group. Finally, using these variables at the airport group level, we employ two different methods, cluster analysis and anomaly detection to identify typologies of disrupted days.

### A. Airport Grouping

There are 496 airports with at least one scheduled commercial flight record in our analysis period. We group these airports based on their Air Route Traffic Control Center (ARTCC) [15], which manages air traffic in a specific region of airspace, and geographic locations. The geographic features and locations of ARTCCs are relatively stable since 2010. Moreover, while commercial service may fluctuate at specific airports, commercial service in a regional sense remains more stable. Therefore, airport groups classified based on these two criteria capture geographic features make their days of operation comparable in the analysis period, which facilities identifying disrupted days.

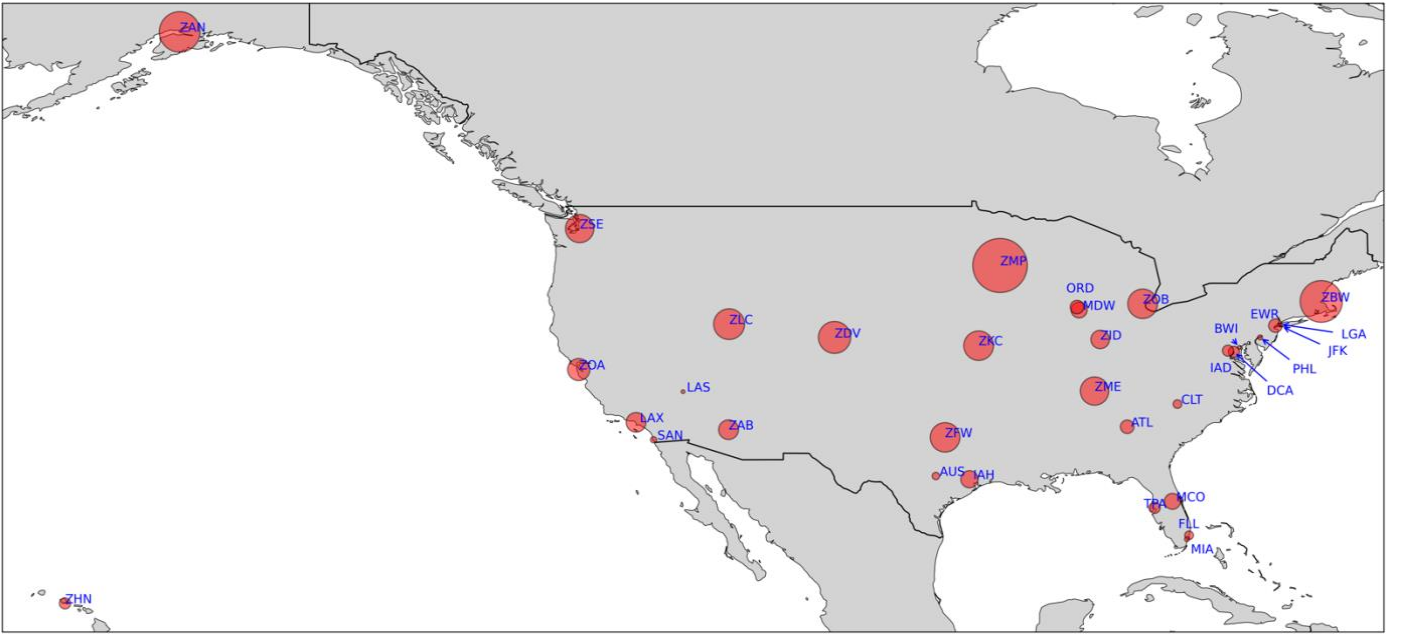

Figure 2. Map of Airport Groups (Size Representing Number of Airports in a Group)

We classify airports into 34 airport groups based on the following algorithm, beginning with a single ARTCC:

1. If there are 0-1 large hub airports within the ATRCC, all airports encompassed in its geographical area define a single airport group. (14 out of 22 ARTCCs)

2. If there are >1 large hub airports within the ATRCC, each large hub airport within the ATRCC becomes an airport group. All non-large hub airports are assigned to an airport group based on their proximity to a large hub airport. (8 out of 22 ARTCCs)

The distribution of airport groups is shown in Figure 2.

*B. Feature Selection*

In order to describe NAS performance with geographical patterns, we select and develop metrics to capture operational performance in airport groups.

Cancellations and delays are direct and important indicators for operational disruptions for air traffic system [14]. Delays include departure delay (DD), airborne delay (AirD), and arrival delay (ArrD); all which measure flight on-time performance at different flight stages.

Departure delay captures pre-departure delays which are cheaper and safer compared to airborne and arrival delay.

$$DD = \max(0, \text{Actual} - \text{Scheduled Departure Time}) \quad (1)$$

Airborne delay captures en-route and pre-landing delays and are relatively more dangerous and expensive.

$$AirD = \max(0, \text{Actual} - \text{Flight Plan Airborne Time}) \quad (2)$$

Arrival delay is the total delay for a flight, accumulating gate delays, airborne delays and taxi delays.

$$ArrD = \max(0, \text{Actual} - \text{Scheduled Arrival Time}) \quad (3)$$

In this study, unit of observation is a day of operations in the NAS. We define "day" of flights as their scheduled departure dates at local times (note that red eye flights scheduled to arrive in the next day are viewed as components of its departure days of operations).

Therefore, we develop aggregated delay and cancellation metrics for days of operations for airport groups. These metrics include cancellation rate, average departure delay, average arrival delay, and average airborne delay. For each airport group ($ag$), we denote daily total number of scheduled flights arriving at the airport group as $A_{ag,day}$, daily total number of scheduled flights departing from the airport group as $D_{ag,day}$, daily number of cancelled arrival flights as $CA_{ag,day}$, and daily number of cancelled departure flights as $CD_{ag,day}$. The metrics for each day of operation for each airport group are defined as follows:

$$CX_{ag,day} = \frac{CA_{ag,day} + CD_{ag,day}}{A_{ag,day} + D_{ag,day}} \quad (4)$$

$$DD_{ag,day} = \frac{\sum_{i=1}^{D_{ag,day}} DD_i}{D_{ag,day}} \quad (5)$$

$$ArrD_{ag,day} = \frac{\sum_{i=1}^{A_{ag,day}} ArrD_i}{A_{ag,day}} \quad (6)$$

$$AirD_{ag,day} = \frac{\sum_{i=1}^{A_{ag,day}} AirD_i}{A_{ag,day}} \quad (7)$$

Ultimately, we have 4 * 34 = 136 features (i.e., 4 daily features for each airport group * 34 airport groups) for each day of operations.

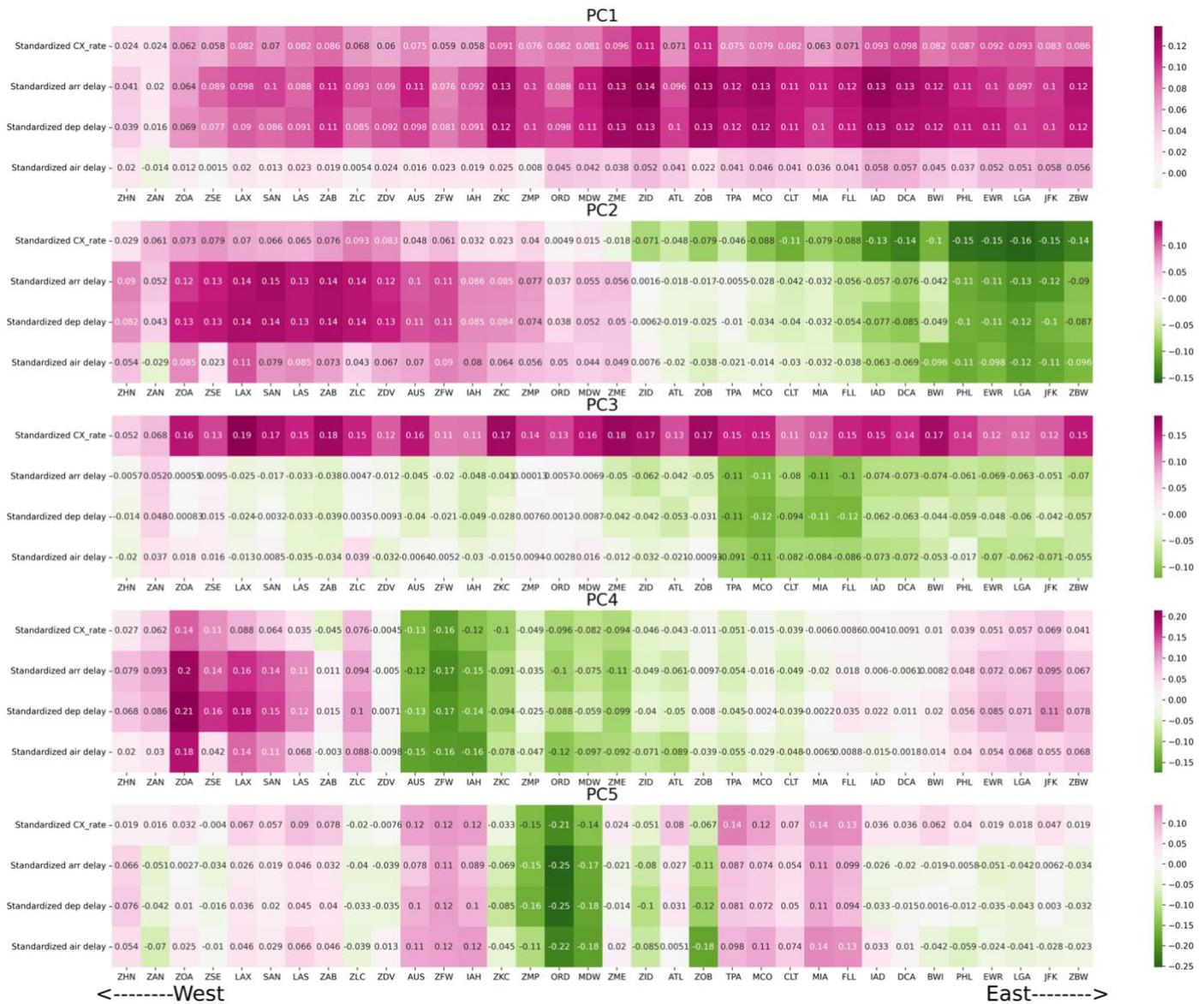

Figure 3. Loadings of Representative Principal Components

## C. Identify Disrupted Days

The result of the feature selection is a 136-dimentional vector capturing daily NAS operational performance. Now we employ two different methods, cluster analysis (i.e., K-Means Cluster Model) and anomaly detection (i.e., Isolation Forest Model) on these data to identify disrupted days of operations with the common set of data and features.

### 1) K-Means Cluster Model

Cluster methods are typically used to cluster similar objects; in doing so, cluster analysis is also an effective way to identify outliers of NAS days of operations. Cluster analysis takes in variables and finds groups that minimize the difference between actual values and a cluster centroid.

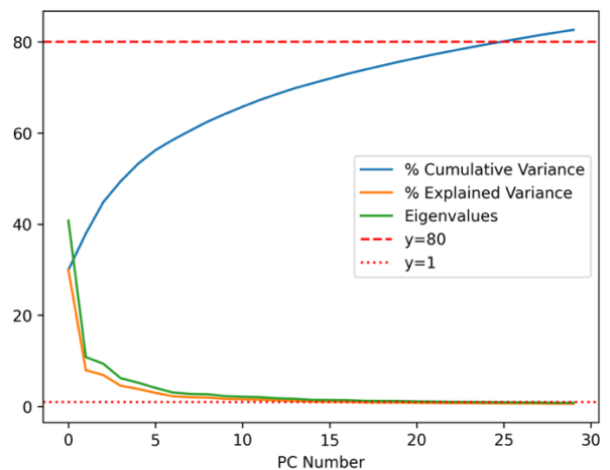

Figure 4. Explained Variance for Different PC Numbers

To facilitate clustering, we employ Principal Component Analysis (PCA) [16] to reduce dimensionality of daily features. PCA works by transforming a dataset into a new coordinate system, and the new coordinate system reduces dimensionality of the original data while preserving essential patterns. In detail, PCA begins with the standardization of the dataset. Next, the covariance matrix is computed to capture the relationships between the variables. Eigenvalues and eigenvectors are then derived from the covariance matrix, with the eigenvalues indicating the variance explained by each principal component. Finally, the data is projected onto the new feature space defined by the selected principal components. In this process, we need to select suitable principal components to reduce dimensionality of original data while retaining the most significant patterns.

For the 136-dimentional dataset representing daily performance of the NAS, we employ PCA to transform it to a new reduced dimensional coordinate system with most variance explained by the principal components. The variance explained by principal components and their eigenvalues are shown in Figure 4. We prefer principal components with eigenvalue >1 because such principal component explains more variance than a single standardized original variable. Accordingly, we select 24 principal components with eigenvalues >1, accounting for 78.8% explained variance. Loadings of some of the principal components are shown in Figure 3. Each row captures one performance metric: standardized features ($CX_{ag,day}$, $ArrD_{ag,day}$, $DD_{ag,day}$, and $AirD_{ag,day}$). Each column captures features for one airport group, and is ordered from left to right by their longitudes from west to east. The red color represents bad performance, i.e., higher cancellation rate and more delays, while green represents good performance. As shown in these figures, we find that loadings of PC1, which has 30.0% explained variance, is almost red. It captures NAS level performance, and a larger value of PC1 represents more NAS level cancellations and delays. Similarly, PC3, which shares 6.9% explained variance, is red in cancellation rate for all airport groups, and is close to 0 for delays. So, it captures NAS level cancellation performance. Conversely, remaining PCs sharing smaller explained variance show obvious and different geographical patterns. For instance, a larger value of PC2 represents good performance in east regions and bad performance in west regions, larger value of PC4 captures bad performance especially at west coast areas. These PCs capture air traffic system performance with multiple metrics and geographic patterns. So, we have a 24-dimentional vector to represent each day of operations for the National Airspace System (NAS).

We then employ K-Means Cluster Model to typologize days of operations. The silhouette score measures how similar data points are to their own cluster compared to other clusters. The score ranges from -1 to 1, with higher values indicating well-separated, distinct clusters. Within-Cluster Scatter (WS) score is another metric to evaluate the cluster results, and it measures the compact nature of clusters. Silhouette scores and WS scores for different numbers of clusters are shown in Figure 5; in short, we are looking for the cluster number that maximizes the Silhouette score and minimizes the WS. Although 2 is the optimal number of clusters considering this decision criteria; there are other local optima to consider. Our analysis is focused on the identification of outlier days. Therefore, we select 12 clusters because the Silhouette score is a local maximum while the WS represents a minor inflection point representing an elbow (meaning that the WS was dropping more steeply up to 12 clusters, then appears to even out).

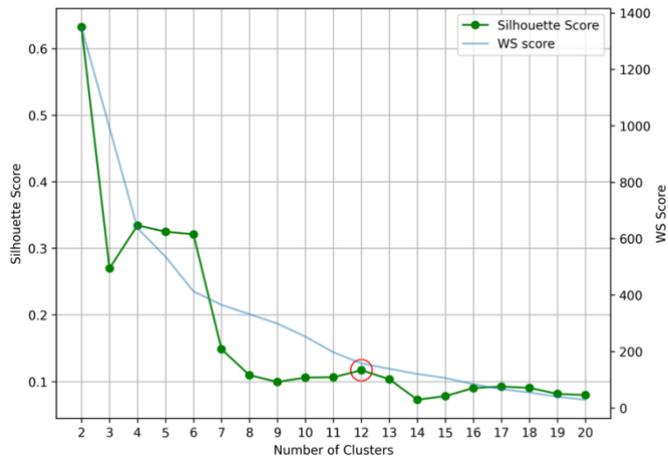

Figure 5. Evaluation of Different Number of Clusters

*2) Isolation Forest Model*

Isolation Forest Model (iForest) [17] is a decision tree-based model designed specifically for anomaly detection. The model is based on an ensemble of trees constructed by randomly choosing a succession of variables and splitting values. In the context of identifying disrupted days of operations, our goal is to identify anomaly performance days—days which are "few and different".

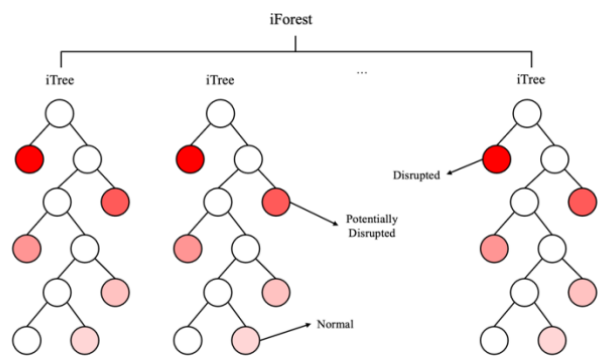

Figure 6. Structure of Isolation Forest

The iForest method as applied to days in the NAS works like this: Because of anomalies' susceptibility to isolation, points of disrupted days will be isolated closer to the root of the trees due to their uniqueness; whereas points of normal days are isolated at the deeper end of the trees due to common features shared with other days as shown in Figure 6 of [17]. Therefore, we use average path length to one particular point in the iForest trees to quantify the level of anomaly of that point. Short average path length indicates anomaly points, which are disrupted days in the study. The average path length is then normalized; values close to 1 indicate anomalies and values closer to 0 indicate normal points. This metric is named as anomaly score.

Figure 7. Aggregated Characteristics of Twelve Clusters

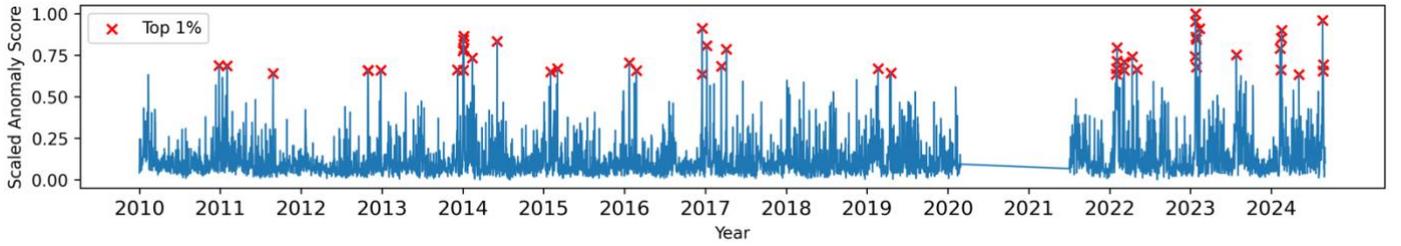

Figure 8. Scaled Anomaly Score for Days of Operations in Analysis Period

TABLE I. CHARACTERISTICS FOR TWELVE CLUSTERS

| Type | Cluster | Cluster Concentration Score (Mean distance to centroid) | Fraction of Days % | Avg Anomaly Scores 0 – Normal 1 – Anomaly | Avg # scheduled flights (CX + operated) | Avg CX rate % | Avg Arrival delay/flight (minutes) |
|---|---|---|---|---|---|---|---|
| Smooth | Smooth1 | 4.10 | 34.42 | 0.06 | 17751 | 1.35 | 7.73 |
| | Smooth2 | 5.85 | 21.71 | 0.07 | 18382 | 2.15 | 12.65 |
| | Almost Smooth | 5.60 | 9.87 | 0.08 | 18344 | 1.91 | 11.29 |
| | East Slight Disturbance | 6.37 | 9.10 | 0.12 | 18686 | 3.52 | 14.60 |
| | NAS Slight Disturbance | 7.12 | 9.29 | 0.15 | 18867 | 3.07 | 19.28 |
| Regional Disturbance | DFW Disturbance | 9.54 | 3.58 | 0.21 | 18852 | 6.26 | 20.70 |
| | ORD Disturbance | 9.54 | 3.43 | 0.23 | 18849 | 7.52 | 21.66 |
| | West Disturbance | 11.78 | 2.87 | 0.24 | 18297 | 4.71 | 18.43 |
| Regional Disruption | Southeast Disruption | 10.94 | 2.30 | 0.30 | 18675 | 4.11 | 23.95 |
| | Northeast Disruption | 9.15 | 5.26 | 0.35 | 19447 | 7.40 | 25.17 |
| NAS Disruption | East Super Disruption | 16.30 | 1.19 | 0.50 | 19332 | 25.29 | 14.19 |
| | NAS Disruption | 21.76 | 0.76 | 0.75 | 18287 | 20.40 | 41.74 |

In this study, we develop an iForest with 100 trees to identify disrupted days. For 4,869 days in our analysis period, each day receives its anomaly score, and high score days are viewed as disrupted days. Disrupted days are the minority consisting of fewer days with poor operational performance in the National Airspace System (NAS), and they have attribute values that are very different from normal days with relatively smooth operations.

IV. RESULTS AND DISCUSSIONS

A. Model Results

In the cluster analysis, we have 12 clusters with their characteristics defined by cluster centroids. Table I presents metrics for the 12 clusters.

In order to identify those clusters that encompass disrupted days, we qualitatively evaluate each cluster. To do so, we visualize each cluster by plotting the cancellation rate on the y-axis vs. the longitude of the airport group on the x-axis; each airport group point is shaded based average delay (Figure 7). These graphs allow us to characterize each cluster by their level of disruption (Smooth, Disturbance, and Disruption) as well as by the geographical nature of the disruption. We further group each individual cluster into four major types according to their level of disruption and geographic characteristics.

The narrative that emerges is:

**Smooth Clusters (5/12 clusters)**: The clusters in this category are two clusters showing smooth, clear days in the NAS (which we call Smooth1 and Smooth2); and three clusters with minor issues but overall low delays and average cancellation rates of less than 4% (which we call Almost Smooth, East Slight Disturbance, and NAS Slight Disturbance). Overall, clusters in the Smooth Category encompass most days in the NAS, and their mean distance to cluster centroid are below 7.5, which are smallest across 12 clusters.

**Regional Disturbance Clusters (3/12 clusters):** Regional Disturbance type clusters represent days with smooth operations in most airport groups but some disturbance in limited number of airport groups. While most airports in each of the three clusters in this group are operating smoothly, each cluster is marked by a limited number of airports that are disrupted, with longer delays and higher cancellation rates; these delays and cancellations do not propagate to their adjacent airport groups.

The names of the three clusters in this group are DFW Disturbance, ORD Disturbance, and West Disturbance.

**Regional Disruption Clusters (2/12 clusters):** The differentiator between the clusters in the Regional Disturbance and the Regional Disruption groups is the level of propagation of delay and cancellation to neighboring airports; this is a marked trait for the cluster in the Regional Disruption category. The clusters named the Southeast Disruption and Northeast Disruption include days with smooth operations in most airport groups but disruptions in other airport groups that have propagated to adjacent airport groups.

**NAS Disruption Clusters (2/12 clusters):** Clusters in the NAS Disruption group – East Super Disruption and NAS Disruption – accounting for 0.8% and 1.2% of the days respectively, are the most disrupted days of operations. The East Super Disruption shows a clear geographic pattern of cancellations with the highest rates at airports in the east. In the NAS Disruption cluster almost all airport groups face long delays and high cancellation rates; in short, the entire air traffic system is under stress. The East Super Disruption days have very high cancellation rates but moderate delays, while both cancellation rates and delays are high in NAS Disruption days.

By comparing four types of clusters, we find that the cluster concentration scores (mean distance to centroid) are lower for Smooth Clusters and higher for NAS Disruption Clusters. Also, Smooth Clusters most days while NAS Disruption Clusters account only 2% of days. It indicates that disrupted days are "few and different", which is consistent with the assumption for the iForest model.

In the iForest model, each day of operation receives its anomaly score ranging between 0 and 1; values closer to 1 indicate an anomaly. Figure 8 shows the time-series anomaly scores and selected features for days in our analysis period. The most disrupted day with the highest anomaly score (=1) is December 23rd, 2022, which is one day in Southwest scheduling crisis. The median anomaly score is 0.09, 95th percentile is 0.37, and 99th percentile is 0.63. The distribution of anomaly score indicates that normal days are similar to each other; while disrupted days are quite different and less than 5% of days might be extremely disrupted and distinguished from normal days with high anomaly scores.

### B. Model Comparison

As we employ two methods to identify disrupted days of operational in the National Airspace System (NAS) with a common set of data and features, comparing these two methods helps validate these new approaches we demonstrate.

#### 1) Areas of Agreement

To evaluate the relationship, the anomaly score of each day within each cluster was recorded. Table I compares average anomaly scores and delay and cancellation metrics for the 12 clusters, and Figure 9 shows the Cumulative Distribution Function (CDF) of all anomaly scores shaded based on its cluster membership. Notice that the portion of the CDF with low anomaly scores corresponds with the Smooth Cluster, as their average anomaly scores are below 0.15. The portion of the CDF with medium anomaly scores corresponds with points in the Regional Disturbance and Regional Disruption clusters. The portion of the CDF with the highest anomaly scores, with average anomaly scores between 0.50 and 0.75, corresponds to those in the NAS Disruption cluster (East Super Disruption and NAS Disruption respectively).

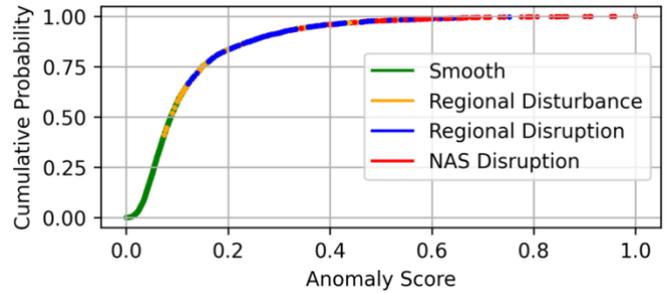

Figure 9: Colored CDF of Anomaly Score

#### 2) Areas of "Disagreement"

A detailed data analysis shows that top anomaly days with high anomaly scores are not always clustered into most disrupted clusters. Figure 10 shows boxplots of anomaly scores as a function of the 12 clusters, ordered from the most smooth to the most disrupted.

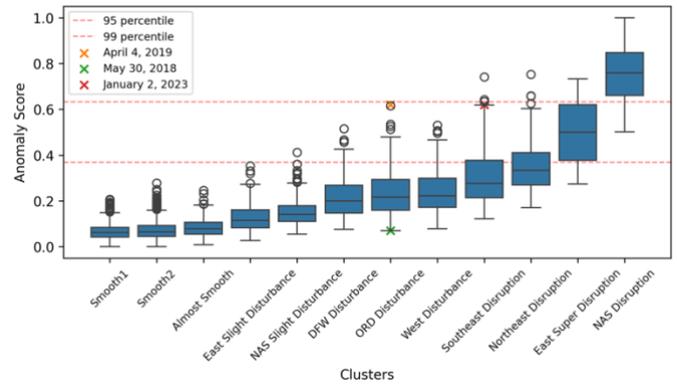

Figure 10. Boxplots of Anomaly Score for Twelve Clusters

TABLE II. COMPARISON OF TWO MODELS

|  | **Cluster Analysis** | **Anomaly Detection** |
| --- | --- | --- |
| Advantages | Based on geographic features. Capturing disruptions with different geographical patterns. | Based on geographic features. Distinguishing severity of disruptions in different number of airport groups. |
| Disadvantages | Only 12 clusters for all days of operations in the long analysis period. For days in the same cluster, failing to further distinguish them. | Failing to capture disruptive geographic patterns. Similar scores for extreme disruptions in few airport groups and moderate disruptions in large amounts of airport groups. |

Although there is generally good agreement in clusters and anomaly scores, some outliers in relatively smooth operated clusters attracts attentions. Even in East Slight Disturbance cluster, which is viewed as Smooth type, there are some days of operations with anomaly score greater than the 95th percentile. This is because days in this cluster may have slightly higher delays and cancellations in most airport groups or just extreme outliers in one or two airport groups. The anomaly detection method is sensitive to these differences, while these differences are not huge enough to distinguish them from other days in their cluster. We further compare the advantages and disadvantages of these two methods for identifying disrupted days in Table II.

By comparing the advantages and disadvantages, we find that each method compensates for the limitations of the other. Therefore, we are able to take the advantages of the two methods and captures disrupted days of operations with not only geographical patterns but also severity of the disruptions. As shown in Figure 11, for instance, April $4^{th}$, 2019, and May $30^{th}$, 2018, are both in ORD Disruption, but their anomaly scores are 0.62 and 0.07 respectively. We can infer that they share the same disruptive patterns with disruptions mainly at ORD area, but the level of disruptions on April $4^{th}$, 2019, is more severe than May $30^{th}$, 2018. In Figure 11, the first two maps visualize the differences in average delays and cancellation rates in color and size of the circles. The disruptions are more severe on April $4^{th}$, 2019, and arrival delays are high in not only ORD area but also in adjacent airport groups. Another contrast pair is April $4^{th}$, 2019, and January $2^{nd}$, 2023. Their anomaly scores are both 0.62, which indicates almost the same severity of disruptions. However, they are clustered in ORD Disruption and Southeast Disruption respectively and share different geographic patterns. In the first map of Figure 11, the disruptions (red large circles) are mostly surrounding ORD area located in the middle part on April $4^{th}$, 2019. While on January $2^{nd}$, 2023, the disruptions are mostly at east coast, especially at southeast part.

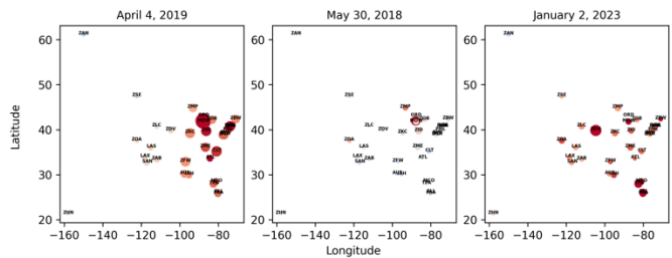

Figure 11: Maps for Representative Days

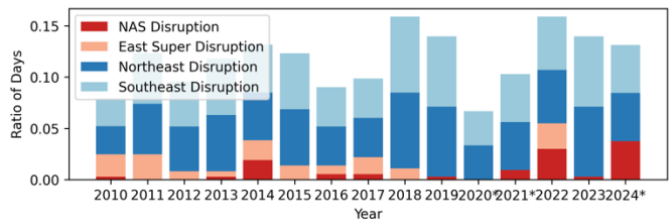

Figure 12: Ratio of Days in Disrupted Clusters for Year

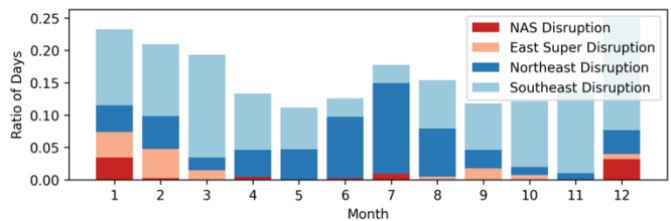

Figure 13: Ratio of Days in Disrupted Clusters for Month

## C. Trends of Disruptions

Clustering and assigning anomaly scores to days of operations back to 2010 provides a chance to investigate the trends in disruptions.

To do so we focus on specific clusters (not the 4 overarching cluster groups). Figure 12 visualizes ratio of number of days in a year in the four most disrupted clusters, i.e., NAS Disruption, East Super Disruption, Northeast Disruption, and Southeast Disruption. Disruptions are more frequent in the years just prior to and after the COVID period.

We then visualize ratio of number of days in four most disrupted clusters for 12 months in Figure 13. In doing so, we can see how the types of disruptions appear through the year. NAS disruptions appear in the winter, summer, and spring – possibly due to their multiple possible causes. Northeast disruptions tend to appear in the spring and early summer (referred to as "SWAP season"); while the fall and winter are heavily dominated by Southeast Disruptions. NAS Disruptions and East Super Disruptions are most frequence in winter, possibly due to snowstorms and heavy traffic in holiday seasons

## D. Delays and Cancellations in Disrupted Days

Finally, we visualize key disruption metrics by cluster towards validating our understanding of the clusters, and how they indicate the severity of delays. We include six features in our analysis period for different clusters, and they are number of days, number of cancellations, OPSNET delays, arrival delays, departure delays, and airborne delays. OPSNET delays are total minutes of System Impact Delay for all delayed operations [18]. For each cluster, we calculate their shares for six features out of all clusters, and then cumulatively them by ordering from Disruption to Smooth. We visualize the cumulative delays and cancellations in different clusters in Figure 14 to investigate the components of delays and cancellations in disrupted days.

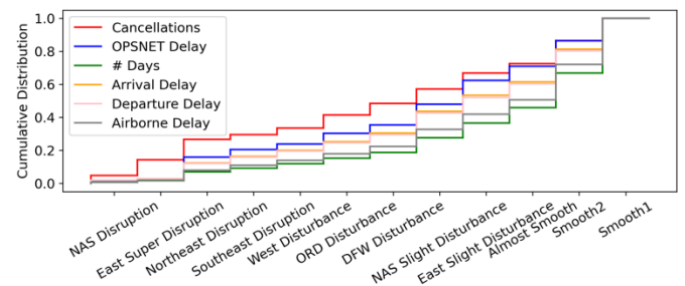

Figure 14. Cumulative Distribution of Selected Metrics in Twelve Clusters

Considering the number of days per cluster, disrupted clusters account for few days across the analysis period: less than 5% for NAS Disruption Clusters, while Smooth Clusters are common accounting for more than 60% of the days. Regarding cancellations, almost 20% of cancellations happened in days in NAS Disruption Clusters. This indicates that cancellations are far less common during normal days of operations. Even in regional disturbance and disruptions, cancellations are not too frequent. OPSNET delay, the blue line, shows a similar, though somewhat less pronounced, pattern, indicating that OPSNET delays occur disproportionately in disrupted clusters but the imbalance is less than for cancellations. Lines of arrival delays and departure delays are in the middle, which indicates that they exist in smooth days and increase in disrupted days. Gray line representing airborne delays is the second bottom line. It indicates that airborne delays are less concentrated in disrupted days, reflecting that such delays, being dangerous and expensive, are mitigated through traffic management even during disruptions.

## V. Conclusions

We presented a rigorous methodology to define typologies of disruptions in the NAS through a combination of cluster analysis and anomaly detection. In doing so, we make the following contributions, some in methodological approach to modeling disruptions and others in applied understanding of disruption.

**First,** we employ cluster analysis (K-Means Cluster Method) and anomaly detection (Isolation Forest Method) to identify disrupted days of operations with consideration of cancellations and delays as well as geographical patterns for all scheduled commercial flights since 2010. The cluster analysis showed a robustness in identifying disrupted days, which we validated with the anomaly score and also data-driven trends for each cluster. Results of the two models show good agreement in localizing disrupted days.

**Second**, the anomaly score method allowed us to further distinguish severity of disruptions for days in the same cluster, and clusters help characterize disruptions' geographical patterns for days with similar anomaly scores. Although days of operations with highest anomaly scores are not all in the most disrupted clusters, the "disagreements" of the two methods compensate for the limitations of each other.

**Third,** we are now able to definitively identify days in the NAS that were disrupted historically. For the results, days in the disrupted clusters accounts for the least percentage, 0.76% for NAS Disruption and 1.19% for East Super Disruption. Days with smooth operations account for highest percentage, as over 85% of the days are in Smooth and Regional Disturbance Type. Also, in anomaly detection results, 95 percentile anomaly score is 0.37, which indicates most days have relatively low (represents normal) scores.

**Fourth,** we were able to further the understanding of the trend of increasing disruptions. The ratio of disrupted days in a year since 2010 is increasing. We see that NAS Disruptions, as well as other clusters of severe disruptions, happened most frequently in the years just prior to and post-COVID. We also identified seasonal patterns between the Northeast and the Southeast, possibly indicating needs for different "playbooks" for these regions.

This methodology developed herein will serve as the foundation for research on countermeasures to disruptions. Because we are able to isolate the days with disruptions and their general typologies and characteristics, it positions us to be able to understand the causes of disruptions and identify mitigation options ultimately increasing the resilience of the NAS in the future. Moreover, now that every day in the NAS historically can be grouped into a cluster, we set the stage for analysis of specific disruptions and their recovery trajectories; in doing so, we can map how certain countermeasures and mitigations assisted in disruption recovery.

## Acknowledgment

Authors thank Alexander S. Estes, Patty Clark, David Lovell, Max Z. Li, and John Schade for their feedback and suggestions.